\documentclass[oneside]{article}
\usepackage{geometry}
\usepackage{graphicx}
\usepackage{amssymb}
\usepackage{latexsym}
\usepackage{amsmath}
\usepackage[toc,page]{appendix}
\usepackage{longtable}
\usepackage{listings}
\usepackage[breaklinks=true]{hyperref}
\usepackage{breakcites}
\usepackage[T1]{fontenc}
\usepackage[english]{babel}
\usepackage{listings}
\usepackage{afterpage}

\usepackage{url}
\date{}

\usepackage{titlesec} 
\renewcommand\thesection{\Roman{section}} 
\renewcommand\thesubsection{\Roman{subsection}} 
\titleformat{\section}[block]{\large\scshape\centering}{\thesection.}{1em}{} 
\titleformat{\subsection}[block]{\large}{\thesubsection.}{1em}{} 

\title{\vspace{-15mm}\fontsize{24pt}{10pt}\selectfont\textbf{Theoretical monitoring of energy transport on solid surfaces at nano-metric scales}} 

\author{
\large
\textsc{Akande Raphael O.}\thanks{\normalsize \href{mailto:raphaelakande@myashiritycode.com}{raphaelakande@myashiritycode.com}},\qquad
\textsc{Oyewande Emmanuel O.}\thanks{\normalsize \href{mailto:oe.oyewande@mail1.ui.edu.ng}{oe.oyewande@mail1.ui.edu.ng}}\\
\normalsize Theoretical Physics Group, University of Ibadan \\ 
\vspace{-5mm}
}

\date{}

\begin{document}

\maketitle 


\begin{abstract}

\noindent 
\textbf{The surface is known to intercept energy from bombarding particles. This energy then spreads over the surface. Before now, it has always been said that the distribution of this energy landing on the surface is always Gaussian. However, in this paper, we clearly show, using a set of mathematical tools, the energy distribution patterns on common, simple or ideal, solid surfaces. We consider flat graphene, cubic and rhombohedra surfaces and indicate the energy leads which transport energy units from one atom to the other, away from the landing site of the bombarding particle. The overall nano-scale pattern of the entire energy spread on the surfaces suggests a clearly non-Gaussian form at nano scales. This means the energy distribution on these surfaces can not be assumed to be uniformly distributed over the surface, at nano scales. The energy travels faster along the length than along the breadth, thus the energy distribution is anisotropic even on ideal lattice, at nano scales. The different patterns, obtained, clearly show that the energy distribution into a material, via its surface, is peculiar to the surface}

\end{abstract}


\providecommand{\keywords}[1]{\textbf{keywords~:~}#1}


\section{Introduction}
Scientific reports, in areas such as nanotherapeutics\cite{1} and surface sputtering, presume a Gaussian distribution on the nano scale range of energy transport. This is not so for all cases, at nano scale, as we show in this paper. The applications of this study are equally suitable for the detection of nano energy shifts from one biological system to other in radiotherapy and micro/nano dosimetry \cite{2}\cite{3} but we shall focus on solid surfaces. Also the maximisation of the activities of catalysts in chemical and biochemical reactions\cite{4} and in the developement of artificial fire resistant materials \cite{5}. Many atimes there are different degrees of energy absorption on different surfaces. Biological surfaces' energies spread differ from that of the inorganics \cite{6}. This may be due to many factors such as photosynthesis and the likes. For instance, in nanotherapeutics, applications of nano particles, bearing nano energy distribution, to biological systems is a good application of this research. It is also important that, in drug development, there is a need to study nano scale energy transport so as to effectively monitor and ensure the delivery of active ingredients and then limit side effects. It is our belief that if the nano energy transport is effectively monitored, as we try to achieve in this paper, much of side effects will be avoided\cite{7}. The metallurgy industry also would require a study of this nano scale energy transport. The surface is a medium and so it is bound to interact with other media and it is in constant tune with its environment. Due to this interaction, there is a need to master the path of the energy (as given off by the foreign particles interacting with it) in form of a wave through the surface \cite{8}. Earlier mathematical models to do this, in the field of surface sputtering, is the Sigmund's Gaussian method \cite{9} and many other models \cite{10},\cite{11}. The model is given as:
\begin{equation}E\left(r'\right)=\frac{\epsilon}{\left(2\pi\right)^{3/2}ab^2 }exp\left(-\frac{z'^2}{2a^2}-\frac{x'^2+y'^2}{2b^2}\right)\end{equation}

where $z'$ is the distance measured along the ion trajectory and $x'$ and $y'$ are measured in the plane perpendicular to $z'$. $\epsilon$ is the total energy carried by the foreign ion (interacting with the surface) and $a$ and $b$ are the widths of the distribution in the directions parallel and perpendicular to the incoming beams respectively. The energy distribution is Gaussian if $a=b$ and not if $a\ne b$. In this paper, we found that $a\ne b$ \cite{12}.


\section{Methods}

To study this energy transport, we want to relate the following, listed, processes to one another. The reason being that they are related in sequential manner. It is our belief that these processes are needed in the energy transport caused by sputtering process.
\begin{enumerate}
\item the location of the landing ions will be called a sputtered location,
\item the origin of the nano-scale energy transport is from the, delivered, energy spread at the sputtered location,
\item all atoms captured by the energy spread will be regarded as the captured atoms,
\item it is important to note that there could be different atomic elements within the captured space,
\item the transportation of this energy causes ionizations of the captured atoms. Different atomic elements will receive and respond to this energy differently,
\item  We plan to apply the, rapidly changing, PAPCs (Photon Absorption Potential Coefficient) of each atom in that captured space to study the variations in the way each atom \textbf{\emph{receives}} energy, 
\item we have to note that even the same atoms, but located at different positions, in the sputtered location would become different due to PAPCs,
\item then we provide a way to evaluate the energy of the captured atoms in the, growing, sputtered location.
\end{enumerate}
To address this issue, we introduce our many-body equation. This equation which is similar to \cite{13} have been developed with the consideration of the facts above. Now, labelling the delivered energy as $E_s$ and the subsequent energies received by the groups of atoms in the, growing, sputtered locations as $E_{ji}$ we have:
\begin{enumerate}
\item 
To get the equation for the first set of atoms to receive energy from a sputtered location, we have:
\allowdisplaybreaks
\begin{eqnarray}E_{q_{11}}=E_s-E_s A_{r_{q11}}=E_s\left(1-A_{r_{q11}}\right),\nonumber\\
E_{q_{12}}=E_s-E_s A_{r_{q12}}=E_s\left(1-A_{r_{q12}}\right) \ldots\nonumber
\end{eqnarray}
So we have all the first categories to be:
\begin{eqnarray}E_{1i}=E_s\left(1-A_{r_{q11}}\right)+E_s\left(1-A_{r_{q12}}\right)+\ldots\nonumber\\
=E_s\left[\left(1-A_{r_{q11}}\right)+\left(1-A_{r_{q12}}\right)+\ldots\right]\nonumber\\
=E_s\left[N-A_{r_{q11}}-A_{r_{q12}}-\ldots-A_{r_{q1N}}\right]\nonumber\\ 
=E_s\sum^{n_i}_i\left(1-A_{r_{q1i}}\right)\end{eqnarray}

\item Similarly, to get the equation for the second set of atoms to receive energy, one of the second set of atoms will have:
\allowdisplaybreaks
\begin{gather*}E_{q_{21}}=E_s-E_s A_{r_{q21}}+E_s\sum^{n_1}_i\left(1-A_{r_{q1i}}\right)-
E_s\sum^{n_1}_i\left(1-A_{r_{q1i}}\right)A_{r_{q_{21}}}\nonumber\\
=E_s\left(1-A_{r_{q21}}\right)+E_s\sum^{n_1}_i\left(1-A_{r_{q1i}}\right)\times 
\left(1-A_{r_{q21}}\right) \nonumber\\
=E_s\left(1-A_{r_{q21}}\right)\Bigg\{1+\sum^{n_1}_i\left(1-A_{r_{q1i}}\right)\Bigg\} \nonumber\\
=E_s\Bigg\{1+\sum^{n_1}_i\left(1-A_{r_{q1i}}\right)-\sum^{n_1}_i\left(1-A_{r_{q1i}}\right)\times 
A_{r_{q21}}-A_{r_{q21}}\Bigg\} \nonumber\\
=E_s\Bigg\{1+\sum^{n_1}_i\left(1-A_{r_{q1i}}\right)\left(1-A_{r_{q21}}\right)-A_{r_{q21}}\Bigg\} 
\end{gather*}

So we have all the second categories to be:
\allowdisplaybreaks
\begin{eqnarray}E_{2i}=E_s\Bigg\{\sum^{n_j}_j\left(1-A_{r_{q2j}}\right)+\sum^{n_i}_i\left(1-A_{r_{q1i}}\right)\times 
\sum^{n_j}_j\left(1-A_{r_{q2j}}\right)\Bigg\}\nonumber\\
=E_s\Bigg\{\sum^{n_j}_j\left(1-A_{r_{q2j}}\right)\left[\sum^{n_i}_i\left(1-A_{r_{q1i}}\right)+1\right]\Bigg\}\end{eqnarray}

\item Similarly, to get the equation for the third set of atoms to receive energy, one of the third set of atoms will have:
\allowdisplaybreaks
\begin{gather*} E_{q_{31}}=E_s-E_s A_{r_{q31}}+ E_s\sum^{n_1}_i\left(1-A_{r_{q1i}}\right)-
E_s\sum^{n_1}_i\left(1-A_{r_{q1i}}\right)A_{r_{q_{31}}}+
E_s\Bigg\{\sum^{n_j}_j\left(1-A_{r_{q2j}}\right)\times\nonumber\\
\left[\sum^{n_i}_i\left(1-A_{r_{q1i}}\right)+1\right]\Bigg\}- 
E_s\Bigg\{\sum^{n_j}_j\left(1-A_{r_{q2j}}\right)\times 
\left[\sum^{n_i}_i\left(1-A_{r_{q1i}}\right)+1\right]\Bigg\}A_{r_{q31}}\nonumber\\
=E_s\left(1-A_{r_{q31}}\right)+E_s\sum^{n_1}_i 
\left(1-A_{r_{q1i}}\right)\times 
\left(1-A_{r_{q31}}\right) 
+E_s\Bigg\{\sum^{n_j}_j\left(1-A_{r_{q2j}}\right)
\left[\sum^{n_i}_i\left(1-A_{r_{q1i}}\right)+1\right]\Bigg\}\times \nonumber\\
\left(1-A_{r_{q31}}\right)\nonumber\\ 
=E_s\{\left(1-A_{r_{q31}}\right)\Bigg\{1+\sum^{n_1}_i 
\left(1-A_{r_{q1i}}\right)+ 
\sum^{n_j}_j\left(1-A_{r_{q2j}}\right)\left[\sum^{n_i}_i\left(1-A_{r_{q1i}}\right)+1\right]\Bigg\}\} \nonumber\\
=E_s\{\left(1-A_{r_{q31}}\right)\Bigg\{
\left[\sum^{n_i}_i \left(1-A_{r_{q1i}}\right)+1\right]\times 
\left(1+\sum^{n_j}_j \left(1-A_{r_{q2j}}\right)\right)\Bigg\}\} \end{gather*}

So we have all the third categories to be:
\allowdisplaybreaks
\begin{eqnarray}
E_{3i}=E_s\sum{n_k}_k\left(1-A_{r_{q3k}}\right)+ 
E_s\sum^{n_i}_i\left(1-A_{r_{q1i}}\right)\sum{n_k}_k\left(1-A_{r_{q3k}}\right)+ 
E_s\sum^{n_j}_j\left(1-A_{r_{q2j}}\right)\times \nonumber\\
\Bigg\{\sum{n_i}_i\left(1-A_{r_{q1i}}\right)+1\Bigg\}\sum{n_k}_k\left(1-A_{r_{q3k}}\right)\nonumber\\
=E_s\Bigg\{\sum^{n_k}_k\left(1-A_{r_{3k}}\right) 
\Bigg\{\sum^{n_j}_j\left(1-A_{r_{2j}}\right)\left[\sum^{n_i}_i\left(1-A_{r_{1i}}\right)+1\right] 
+\sum^{n_i}_i\left(1-A_{r_{1i}}\right)+1\Bigg\}\Bigg\}\end{eqnarray}

\item Similarly, to get the equation for the fourth set of atoms to receive energy, one of the fourth set of atoms will have:
\allowdisplaybreaks
\begin{gather*}
E_{q_{41}}=E_s-E_s A_{r_{q41}}+E_s\sum^{n_1}_i\left(1-A_{r_{q1i}}\right)- 
E_s\sum^{n_1}_i\left(1-A_{r_{q1i}}\right)A_{r_{q_{41}}}+
E_s\Bigg\{\sum^{n_j}_j\left(1-A_{r_{q2j}}\right) \nonumber\\
\left[\sum^{n_i}_i\left(1-A_{r_{q1i}}\right)+1\right]\Bigg\}-
E_s\Bigg\{\sum^{n_j}_j\left(1-A_{r_{q2j}}\right) 
\left[\sum^{n_i}_i\left(1-A_{r_{q1i}}\right)+1\right]\Bigg\}A_{r_{q41}}\nonumber\\
+\Bigg\{E_s\Bigg\{\sum^{n_k}_k\left(1-A_{r_{q3k}}\right) 
\Bigg\{ \sum^{n_j}_j\left(1-A_{r_{q2j}}\right)\left[\sum^{n_i}_i\left(1-A_{r_{q1i}}\right)+1\right]+ 
\sum^{n_i}_i\left(1-A_{r_{q1i}}\right)+1 \Bigg\}\Bigg\}\Bigg\}\nonumber\\
-\{E_s\Bigg\{\sum^{n_k}_k\left(1-A_{r_{q3k}}\right)\times 
\Bigg\{\sum^{n_j}_j\left(1-A_{r_{q2j}}\right)\left[\sum^{n_i}_i\left(1-A_{r_{q1i}}\right)+1\right]+ 
\sum^{n_i}_i\left(1-A_{r_{q1i}}\right) +1\Bigg\}\Bigg\}A_{r_{q41}}\}\nonumber\\
=E_s\left(1-A_{r_{q41}}\right)+E_s\sum^{n_1}_i\left(1-A_{r_{q1i}}\right)\times 
\left(1-A_{r_{q41}}\right)+ 
E_s\Bigg\{\sum^{n_j}_j\left(1-A_{r_{q2j}}\right)\left[\sum^{n_i}_i\left(1-A_{r_{q1i}}\right)+1\right]\Bigg\}\times\nonumber\\
\left(1-A_{r_{q41}}\right) + E_s \Bigg\{ \sum^{n_k}_k\left(1-A_{r_{q3k}}\right) \times 
\{ \sum^{n_j}_j\left(1-A_{r_{q2j}}\right) [ \sum^{n_i}_i\left(1-A_{r_{q1i}}\right)+1 ] + 
\sum^{n_i}_i\left(1-A_{r_{q1i}}\right)+1 \}\Bigg\}\left(1-A_{r_{q41}}\right)\end{gather*}
\begin{equation}\end{equation}

So we have all the fourth categories to be:
\allowdisplaybreaks
\begin{gather*}E_{4i}=E_s\sum^{n_l}_l\left(1-A_{r_{q4l}}\right) \Bigg\{ \sum^{n_k}_k\left(1-A_{r_{q3k}}\right)\times 
\Bigg\{ \sum^{n_j}_j\left(1-A_{r_{q2j}}\right)\{\sum^{n_i}_i\left(1-A_{r_{q1i}}\right)+1 \} +\nonumber\\
\sum^{n_i}_i\left(1-A_{r_{q1i}}\right)+1\Bigg\}+ \sum^{n_j}_j\left(1-A_{r_{q2j}}\right) 
\Bigg\{\sum^{n_i}_i\left(1-A_{r_{q1i}}\right)+1\Bigg\}+ 
\sum^{n_i}_i\left(1-A_{r_{q1i}}\right)+1 \Bigg\}\end{gather*}
\begin{equation}\end{equation}

\end{enumerate}

In general, we have:

\begin{gather*}
\allowdisplaybreaks
E_{nm}=E_s \Bigg\{ \sum^{n_m}_m\left(1-A_{r_{nm}}\right) 
\Bigg\{ \sum^{n_{m-1}}_\tau\left(1-A_{r_{n-1,\tau}}\right) \times 
\Bigg\{ \sum^{n_{m-2}}_\eta\left(1-A_{r_{n-2,\eta}}\right)\ldots 
\sum^{n_i}_i\left(1-A_{r_{1i}}\right)+1 \Bigg\} +\nonumber\\
\sum^{n_i}_i\left(1-A_{r_{1i}}\right)+1\Bigg\}+\ldots+ 
\sum^{n_i}_i\left(1-A_{r_{2i}}\right)\times 
\left[\sum^{n_i}_i\left(1-A_{r_{2i}}\right)+1\right]+\sum^{n_i}_i\left(1-A_{r_{1i}}\right)+1\Bigg\} 
\end{gather*}
\begin{equation}\end{equation}

Please note that all limits of the summation signs are mere dummy variables. We shall label equation (7) as the Photon Potential Difference (PPD) which we have derived to compute the consumption of sputtering energy, landing on the sputtering location, by the atoms in the sputtering location region. We plan to employ the electrostatic repulsion of the ionised atoms to explain the reason for the sputtering events. Therefore, we shall carry out our computations by \textbf{\emph{checking}} or \textbf{\emph{validating}} the sum of ionisation energies, $I$, of all the atoms within the, growing, sputtering location. However, we shall replace the $I$ with the rates of PAPCs, $A_r$, of the constituent atoms. Hence, the validity $\varsigma$ of a sputtering event is given as:
\begin{eqnarray}\varsigma=I_{n,m}\le\partial_{r_{nm}}\left(E_{n,m}\right)\le I _{n,\mu}\nonumber\\
\varsigma=A_{r_{n,m}}\le\partial_{r_{nm}}\left(E_{n,m}\right)\le A_{r _{n,\mu}}\end{eqnarray}
Where $\partial_{r_{nm}}\left(E_{n,m}\right)=\frac{d\tilde{v}}{d N_{nr}}\to \frac{E_{ion}}{\tilde{N}\left(A_c\right)}$ is the gradual decay of energy as the sputtering location grows.
where PAPC is given as:
\begin{eqnarray}A_c=\frac{1}{2 z}\left\{\ln\left(\frac{1+z^-}{1-z^-}\right)+\ln\left(\frac{1+z^+}{1-z^+}\right)\right\}\nonumber\\
=\frac{1}{2 z}\left\{\ln\left(\frac{1+z^- +z^+ +z^-\times z^+}{1-z^-  -z^+ +z^-\times z^+}\right)\right\}\end{eqnarray}
Equation(9) can also be written as:
\begin{equation}A_c=\frac{1}{z}\left( \tanh^{-1}(z^-) + \tanh^{-1}(z^+)\right)\end{equation}
where $z^-$ is is given as:
\begin{equation}z^-=\frac{z_v^-}{\left|z-z_v^-\right|}\end{equation}
and $z^+$ is is given as:
\begin{equation}z^+=\frac{z_v^+}{\left|z-z_v^+\right|}\end{equation}
where $z_v^+$ and $z_v^-$ are the extra subtracted and added electrons to the neutral atom in question. Take for instance, the $A_c$ of an atom like $Na$ is will be given as:
\begin{itemize}
\item for first ionisation, we have:
$z_v^- = \frac{12}{\left|11-12\right|}=12$ and $z_v^+ = \frac{10}{\left|11-10\right|}=10$. $A_c=0.0167148$
\item for second ionisation, we have:
$z_v^- = \frac{13}{\left|11-13\right|}=6.5$ and $z_v^+ = \frac{9}{\left|11-9\right|}=4.5$. $A_c=0.0346427$
\item for third ionisation, we have:
$z_v^- = \frac{14}{\left|11-14\right|}=4.66667$ and $z_v^+ = \frac{8}{\left|11-8\right|}=2.66667$. $A_c=0.0556262$ 
\end{itemize}
\begin{eqnarray}A_r=A_c\frac{c}{R}\nonumber\end{eqnarray}
where $c$ is the speed of the transported photon and $R$ is the increasing distance of the transporting energy from the landing site of the foreign particle.

\section{Inteference of growing sputtering location}
Since, equation (7) is for only one sputtered location, therefore, we need to provide ways to handle the interference of more than one sputtering locations that \textbf{\emph{intefere}}. We are not trying to \textbf{\emph{merge}} isolated sputtering locations. What we intend doing is to compute the energy consumption in all \textbf{\emph{interfering}} sputtering locations. Therefore, each non-intefering sputtering locations are treated, and isolated, differently. 

Now, for a sputtering location we have the \textbf{\emph{irregular}} energy flow to be :
\begin{equation}\hat{\varsigma}=\bigcap_{\varsigma\in A_{n,m<\mu}}\int^{n'}_n \int^{m'}_m\partial_{r_{mn}}\left(E_{n,m,\varsigma}\right) dm dn\end{equation}
Equation (13), above, is the extent of energy spread within the validity. The number of captured  atoms, expected to be sputtered, is $E_{n,m,\varsigma}$. It is on these atoms that we will perform the derived potential, given as equation (7). The double integral in equation (13) is to handle the \textbf{\emph{irregular}} flow of energy in the, growing, sputtered location.

Due to different energies from different ions, landing on the surface at different locations and having their sputtering locations merge, the PPD seems to be from different sources. Therefore, the energy reaching the $nm^{th}$ atoms is given as:
\begin{eqnarray}\hat{\varsigma}=\bigcup_{\Gamma_{x,y,z}}\sum^{t_{max}}_{t_o}\bigcup_{t^-\le t\le t^+} \Bigg\{\bigcap_{\varsigma\in A_{n,m<\mu}}\int^{n'}_n \int^{m'}_m 
\partial_{r_{mn}}\left(E_{n,m,\varsigma,t}\right) dm dn\Bigg\}\end{eqnarray}
Where $\bigcup_{t^-\le t\le t^+}$ is for the possible merger of more than one sputtered locations, $t_o$ and $t_{max}$ are the initial and final times of the landing time differences between first ion and the last ion. There could be several ions in between. The quantities $t^-$ and $t^+$ are given as: 
\begin{eqnarray}t^+=t^+_o+\bigtriangleup t, \qquad t^-=t^-_o+\bigtriangleup t \end{eqnarray}
For more than two ions, we have :
\begin{eqnarray}t_o=t^+_o-t^-_o+\bigtriangleup t \qquad\nonumber\\ 
t_1=\left[\left(t^+_o-t^-_o\right)-\left(t^+_1-t^-_1\right)\right]+\bigtriangleup t \qquad\nonumber\\ 
t_2=\Big\{\left(t^+_o-t^-_o\right)-\left(t^+_1-t^-_1\right)-\left(t^+_2-t^-_2\right)\Big\}+\bigtriangleup t \ldots \end{eqnarray}
In general, we shall have the following as the time difference between several landing ions:
\begin{eqnarray}t_n=\Big\{\left(t^+_o-t^-_o\right)-\ldots -\left(t^+_n-t^-_n\right) \Big\}+\bigtriangleup t \end{eqnarray}
The quantities $t^-$ and $t^+$ are designed to help us track the time lapse between any two events. Where $\bigtriangleup t$ is the continuously increasing time step for all landing ions. $t^+_o$ and $t^-_o$ are the initials of any two landing ions and are always increasing. The $+$ and $-$ signs denote the first and next to land respectively. 

The $\sum^{t_{max}}_{t_o}$ sums up all the landing energies as they land according to timing and the energy reaching the $nm^{th}$ atom, i.e $E_{n,m,\varsigma,t}$, now has the time tag to indicate different energies with time. This is because the energy at a first instance would have died down, a bit, before another falls. Therefore, the summation is not the energy that lands but the energy that remains amongst the atoms. The quantity $\bigcup_{\Gamma_{x,y,z}}$ is the parameter that \textbf{\emph{records}} the displaced atoms in the surface for each time frame.  
\newline

It is important to note that this sputtered location undergoes a form of vibration due to movement of atoms to and fro in the lattice. In order to account for this, seeming error, we shall propose that the sputtered location or its constituent atoms behave in a simple harmonic motion manner. Our proposal is based on the fact that the creation of same charge, around the sputtered location, depends on the number of atoms and the $A_c$ of those atoms. \textbf{\emph{It also depends on the frequency with which the atoms vibrate about their $A_c$ values}}. This is what interests us most because, it determines the manner in which the resulting material will react to its environment.

The force of repulsion as the atoms become of the same charge is given as:
\begin{eqnarray}F\propto -\psi\Rightarrow F=-\tau \psi,\qquad \tau=\delta\left(A_c\right)\nonumber\\ 
m\ddot{\psi}=F\Rightarrow m\ddot{\psi}=-\delta\left(A_c\right) \psi\nonumber\\
\ddot{\psi}+\omega^2_o\psi=0 \Rightarrow \omega^2_o=\left|\left(\frac{\delta\left(A_c\right)}{m}\right)^{1/2}\right|\nonumber\\
\psi\left(A_c,t\right)=\phi \cos\left(\omega_o t + \varphi\right)\end{eqnarray}
Equation (18) depends on both $t$ and $A_c$ because both are varying through out the process. We shall let $\phi$ be the geometry of the sputtered location and $\varphi$ be the number of atoms, possible, in that geometry. However, in this work, we shall limit ourselves to the distribution of energy only.


\section{Results}

In this section, we shall present the results of our developed models by considering a much detailed energy transport and sharing by each atoms in the three lattice structures considered. We consider the atoms, on whom an ion lands, to be scattered in every direction. We have chosen three axes with a consideration that the third, $z$, will not scatter energy like the other two: $x$ and $y$. Here goes the model:
\begin{eqnarray}\bigtriangledown\cdot\phi^\lambda=\left(\underline{i}\frac{\partial}{\partial x}+\underline{j}\frac{\partial}{\partial y}+\underline{k}\frac{\partial}{\partial z}\right)\cdot\Big\{\underline{i}a^\lambda_i+\underline{j}b^\lambda_i+ \underline{k}\left[\partial c^{\mu\nu}_i+\partial d^{\mu\nu}_i\right]\Big\}\nonumber\\
\bigtriangledown\cdot\phi^\lambda=\frac{\partial a^{\lambda_x}_i}{\partial \lambda_x}+\frac{\partial b^{\lambda_y}_i}{\partial \lambda_y}+\left(\frac{\partial^2 c^{\mu\nu}_i}{\partial \mu\nu}+\frac{\partial^2 d^{\mu \partial \nu}_i}{\partial \mu \partial \nu}\right)\end{eqnarray}

Where $i$ is for the labeling of the four cardinal points and $\lambda$ is for identifying the sputtered location amongst others. Equation (19) is the energy spread by only one landing ion on a 2D \textbf{flat} surface. The beauty of this equation is that it does not need the angle of landing. Energy just spreads according to the potential describing the landing. Now, for more than one \textbf{\emph{merging}} ion landing sites, on a 2D \textbf{flat} surface, we have:

\begin{eqnarray}\bigtriangledown\cdot\phi^\lambda_j=\bigtriangledown\cdot\left(\phi^\lambda_1+\phi^\lambda_2+\ldots+\phi^\lambda_n\right)
\nonumber\\ 
\bigtriangledown\cdot\phi^\lambda_j=\sum^N_j\Bigg\{\frac{\partial a^{\lambda_x}_{ij}}{\partial \lambda_x}+\frac{\partial b^{\lambda_y}_{ij}}{\partial \lambda_y}+\left(\frac{\partial^2 c^{\mu\nu}_{ij}}{\partial \mu \partial \nu}+\frac{\partial^2 d^{\mu\nu}_{ij}}{\partial \mu \partial \nu}\right)\Bigg\}\end{eqnarray}

Now, for more than one merging landing ions on a rough (with respect to a height variant $\sigma$) 2D flat surface, which we simply refer to as a 3D surface, we have:
\begin{eqnarray}\frac{\partial \bigtriangledown\phi^\lambda_j}{\partial\sigma}=
\frac{\partial\left(\bigtriangledown\left[\phi^\lambda_1+\phi^\lambda_2+\ldots+\phi^\lambda_n\right]\right)}{\partial\sigma}\nonumber\\ \frac{\partial \bigtriangledown\phi^\lambda_j}{\partial\sigma}=\sum^N_j\Bigg\{\frac{\partial^2 a^{\lambda_x}_{ij\sigma}}{\partial \lambda_x \partial \sigma}+\frac{\partial^2 b^{\lambda_y}_{ij\sigma}}{\partial \lambda_y \partial \sigma}+
\left(\frac{\partial^3 c^{\mu\nu}_{ij\sigma}}{\partial \mu \partial \nu \partial \sigma}+\frac{\partial^3 d^{\mu\nu}_{ij\sigma}}{\partial \mu \partial \nu \partial \sigma}\right)\Bigg\}\end{eqnarray}

The interesting thing about this new, advanced, version of our models so far is that it can give us the energy reaching an atom per the type of atom receiving the energy. That is, different atoms placed in the lattice will receive a corresponding energy to its current position. So we have:
\begin{eqnarray}\frac{\partial \bigtriangledown\phi^\lambda_j}{\partial\sigma}=\sum^N_j\Bigg\{\frac{\partial^2 a^{\lambda_x}_{ij\sigma}}{\partial \lambda_x \partial \sigma}A^{\mu\nu}+\frac{\partial^2 b^{\lambda_y}_{ij\sigma}}{\partial \lambda_y \partial \sigma}B^{\mu\nu} +\left(\frac{\partial^3 c^{\mu\nu}_{ij\sigma}}{\partial \mu \partial \nu \partial \sigma}C^{\mu\nu}+\frac{\partial^3 d^{\mu\nu}_{ij\sigma}}{\partial \mu \partial \nu \partial \sigma}D^{\mu\nu}\right)\Bigg\}\end{eqnarray}

where $A^{\mu\nu}$, $B^{\mu\nu}$, $C^{\mu\nu}$ and $D^{\mu\nu}$ are the matrices intended to multiply $\frac{\partial^2 a^{\lambda_x}_{ij\sigma}}{\partial \lambda_x \partial \sigma}$, $\frac{\partial^2 b^{\lambda_y}_{ij\sigma}}{\partial \lambda_y \partial \sigma}$, $\frac{\partial^3 c^{\mu\nu}_{ij\sigma}}{\partial \mu \partial \nu \partial \sigma}$ and $\frac{\partial^3 d^{\mu\nu}_{ij\sigma}}{\partial \mu \partial \nu \partial \sigma}$ respectively so as to determine which atom is recieving what energy. Equation (22) describes, in full details, the entire energy spread and the energy reaching each atom in a \textbf{\emph{cubic lattice}} as those considered in this paper. We thereby conclude that \textbf{\emph{the different patterns, as shown in figures 1 and 6, clearly show that the energy distribution into a material, via its surface, is peculiar to the surface.}} We have:
\allowdisplaybreaks
\begin{eqnarray}
A^{\mu\nu}_1=\left(\begin{array}{cc}K_{n-\left(2\lambda_x+1\right),m,\sigma} & -1 \\K_{n-\left(2\lambda_x+1\right),m+1,\sigma} & 1 \end{array}\right)_{i=1} , \qquad 
A^{\mu\nu}_2=\left(\begin{array}{cc}K_{n,m+\left(2\lambda_x+3\right),\sigma} & -1 \\K_{n+1,m+\left(2\lambda_x+3\right),\sigma} & 1 \end{array}\right)_{i=2}  \nonumber\\
A^{\mu\nu}_3=\left(\begin{array}{cc}K_{n+\left(2\lambda_x+2\right),m+1,\sigma} & -1 \\K_{n+\left(2\lambda_x+2\right),m+2,\sigma} & 1 \end{array}\right)_{i=3}  , \qquad 
A^{\mu\nu}_4=\left(\begin{array}{cc}K_{n,m+2\lambda_x,\sigma} & -1 \\K_{n+1,m+2\lambda_x,\sigma} & 1 \end{array}\right)_{i=4}\nonumber\\
B^{\mu\nu}_1=\left(\begin{array}{cc}K_{n-\lambda_x,m+\left(1-\lambda_x\right),\sigma} & 0 \\0 & 1 \end{array}\right)_{i=1}  , \qquad 
B^{\mu\nu}_2=\left(\begin{array}{cc}K_{n-\lambda_x,m+\left(\lambda_x+2\right),\sigma} & 0 \\0 & 1 \end{array}\right)_{i=2}  \nonumber\\
B^{\mu\nu}_3=\left(\begin{array}{cc}K_{n+\left(\lambda_x+1\right),m+\left(\lambda_x+2\right),\sigma} & 0 \\0 & 1 \end{array}\right)_{i=3}  , \qquad 
B^{\mu\nu}_4=\left(\begin{array}{cc}K_{n+\left(\lambda_x+1\right),m+\left(1-\lambda_x\right),\sigma} & 0 \\0 & 1 \end{array}\right)_{i=4} \nonumber\\
C^{\mu\nu}_1=\left(\begin{array}{cc}K_{n-\left(2\lambda_x+2\right)+\mu,m+\nu,\sigma} & -1 \\K_{n-\left(2\lambda_x+2\right)+\mu,m+\nu,\sigma} & 1 \end{array}\right)_{i=1}  , \qquad 
C^{\mu\nu}_2=\left(\begin{array}{cc}K_{n-\left(2\lambda_x+2\right)+\mu,m+\left(2\lambda_x+2\right)+\nu,\sigma} & -1 \\K_{n-\left(2\lambda_x+2\right)+\mu,m+\left(2\lambda_x+3\right)+\nu,\sigma} & 1 \end{array}\right)_{i=2}  \nonumber\\
C^{\mu\nu}_3=\left(\begin{array}{cc}K_{n+\left(2\lambda_x+3\right)+\mu,m+\left(2\lambda_x+2\right)+\nu,\sigma} & -1 \\K_{n+\left(2\lambda_x+3\right)+\mu,m+\left(2\lambda_x+3\right)+\nu,\sigma} & 1 \end{array}\right)_{i=3}  , \qquad 
C^{\mu\nu}_4=\left(\begin{array}{cc}K_{n+\left(2\lambda_x+3\right)+\mu,m+\nu,\sigma} & -1 \\K_{n+\left(2\lambda_x+3\right)+\mu+1+\nu,m,\sigma} & 1 \end{array}\right)_{i=4} \nonumber\\
D^{\mu\nu}_1=\left(\begin{array}{cc}K_{n-1+\mu,m-\left(2\lambda_x+1\right)+\nu,\sigma} & -1 \\K_{n+\mu,m-\left(2\lambda_x+1\right)+\nu,\sigma} & 1 \end{array}\right)_{i=1}  , \qquad 
D^{\mu\nu}_2=\left(\begin{array}{cc}K_{n-1+\mu,m+\left(2\lambda_x+4\right)+\nu,\sigma} & -1 \\K_{n+\mu,m+\left(2\lambda_x+4\right)+\nu,\sigma} & 1 \end{array}\right)_{i=2}  \nonumber\\
D^{\mu\nu}_3=\left(\begin{array}{cc}K_{n+1+\mu,m+\left(2\lambda_x+4\right)+\nu,\sigma} & -1 \\K_{n+2+\mu,m+\left(2\lambda_x+4\right)+\nu,\sigma} & 1 \end{array}\right)_{i=3}  , \qquad 
D^{\mu\nu}_4=\left(\begin{array}{cc}K_{n+1+\mu,m-\left(2\lambda_x+1\right)+\nu,\sigma} & -1 \\K_{n+2+\mu,m-\left(2\lambda_x+1\right)+\nu,\sigma} & 1 \end{array}\right)_{i=4} \nonumber\\
\end{eqnarray}

We can also have:
\begin{eqnarray}
A^{\mu\nu}=\left(\begin{array}{cc}A^{\mu\nu}_1 & A^{\mu\nu}_2 \\ A^{\mu\nu}_3 & A^{\mu\nu}_4 \end{array}\right)
, \qquad 
B^{\mu\nu}=\left(\begin{array}{cc}B^{\mu\nu}_1 & B^{\mu\nu}_2 \\ B^{\mu\nu}_3 & B^{\mu\nu}_4 \end{array}\right)
\nonumber\\
C^{\mu\nu}=\left(\begin{array}{cc}C^{\mu\nu}_1 & C^{\mu\nu}_2 \\ C^{\mu\nu}_3 & C^{\mu\nu}_4 \end{array}\right)
, \qquad 
D^{\mu\nu}=\left(\begin{array}{cc}D^{\mu\nu}_1 & D^{\mu\nu}_2 \\ D^{\mu\nu}_3 & D^{\mu\nu}_4 \end{array}\right)
\end{eqnarray}

\section{Conclusion}
The whole essence of this work is to, theoretically, monitor or determine the true flow path of nano-scale energies on a graphene surface and other surfaces mentioned earlier. Having done all the derivations, we have achieved our goal. We can simply evaluate the flow of each foreign particle landing on the surface at any time using equations (7) in (22). We now know the energy flowing away from the sputtered location and the energy reaching each atom along the way. The parameter PAPC has helped in the easy computation by providing us with information about how each (different atomic elements) respond to this energy flow. The PAPC, therefore, helps to determine the eventual path of the energy flow. In essence, for all the surfaces considered and other similar surfaces, there will be a shift from the path of energy flow, as shown in the figures 1 to 6. This will be in correspondence to the displacement $c$, in the figures 1 to 6, of the center at which the black dots, symbolising the foreign particle, lands. It is our belief that if the nano energy transport is effectively monitored, both theoretically and experimentally, as we try to achieve in this paper, much of side effects of medical treatments and effects such as this will be avoided. The different patterns, figures 1 and 6, clearly show that the energy distribution into a material, via its surface, is peculiar to the surface.
Now, we present the figures showing the leads (arrows) with which we model the nano-scale energy flow:


\begin{figure*}[htb!]
\includegraphics[width=\textwidth, height=8.56cm]{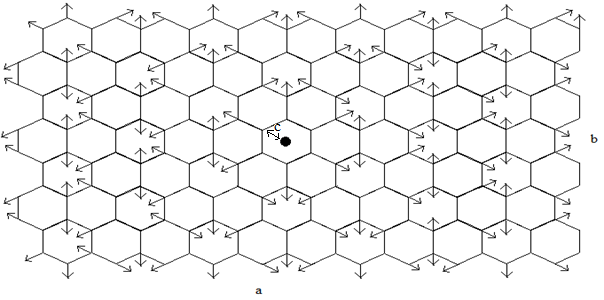}
\caption{ Illustration of an energy distribution from an ion (black dot) landing on an atom (line crossings) on a graphene or hexagonal surface for $c = 0$. The side (b) happens to be the slow end while side (a) is the side with faster energy flow. The displayed pattern will shift in correspondence to $c\ne 0$.}
\end{figure*}

\clearpage

\begin{figure*}[htb!]
\includegraphics[width=\textwidth, height=8.56cm]{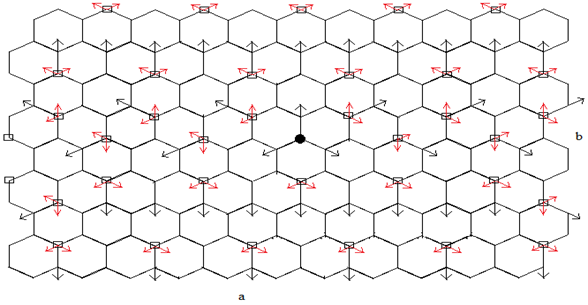}
\caption{ Illustration of an energy distribution from an ion (black dot) landing on an atom (line crossings) on a graphene or hexagonal surface. The side (b) happens to be the slow end while side (a) is the side with faster energy flow. The red lines are the points where the resultant flow is taken.}
\end{figure*}

\begin{figure*}[htb!]
\includegraphics[width=\textwidth, height=8.56cm]{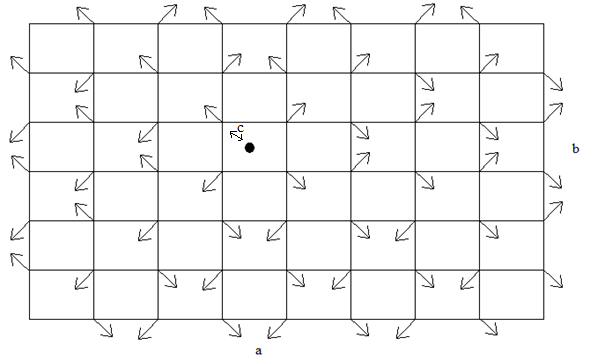}
\caption{ Illustration of an energy distribution from an ion (black dot) landing on an atom (line crossings) on a cubic surface for $c = 0$. The side (b) happens to be the slow end while side (a) is the side with faster energy flow. The displayed pattern will shift in correspondence to $c\ne 0$.}
\end{figure*}

\clearpage

\begin{figure*}[htb!]
\includegraphics[width=\textwidth, height=8.56cm]{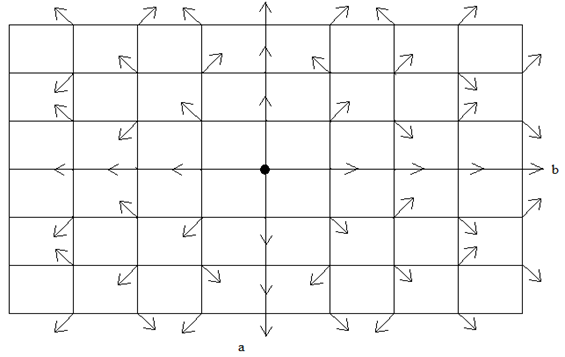}
\caption{ Illustration of an energy distribution from an ion (black dot) landing on an atom (line crossings) on a cubic surface. The side (b) happens to be the slow end while side (a) is the side with faster energy flow. The red lines are the points where the resultant flow is taken.}
\end{figure*}

\begin{figure*}[htb!]
\includegraphics[width=\textwidth, height=8.56cm]{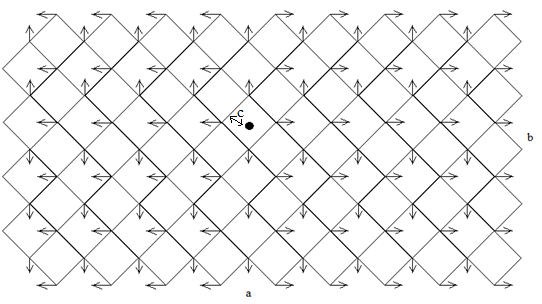}
\caption{ Illustration of an energy distribution from an ion (black dot) landing on an atom (line crossings) on a rhombohedra surface for $c = 0$. The side (b) happens to be the slow end while side (a) is the side with faster energy flow. The displayed pattern will shift in corresppondence to $c\ne 0$.}
\end{figure*}

\clearpage

\begin{figure*}[htb!]
\includegraphics[width=\textwidth, height=8.56cm]{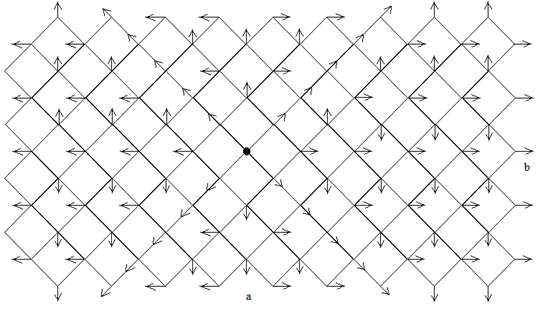}
\caption{ Illustration of an energy distribution from an ion (black dot) landing on an atom (line crossings) on a rhombohedra surface. The side (b) happens to be the slow end while side (a) is the side with faster energy flow. The red lines are the points where the resultant flow is taken.}
\end{figure*}





\end{document}